\documentclass[prl,aps,twocolumn,showpacs,superscriptaddress,floatfix]{revtex4}
\usepackage{graphicx}
\usepackage{bm}
\usepackage{color}
\pdfoutput=1 

\begin{document}
\title{Quantized squeezing and even--odd asymmetry of trapped bosons}

\author{Shijie Hu}
\affiliation{Department of Physics, Renmin University of China, Beijing 
100872, China} 
\affiliation{Institute of Theoretical Physics, CAS, Beijing 100080, China}
\author{Yuchuan Wen}
\affiliation{Institute of Theoretical Physics, CAS, Beijing 100080, China}
\author{Yue Yu}
\affiliation{Institute of Theoretical Physics, CAS, Beijing 100080, China}
\author{B. Normand}
\affiliation{Department of Physics, Renmin University of China, Beijing 
100872, China} 
\author{Xiaoqun Wang}
\affiliation{Department of Physics, Renmin University of China, Beijing 
100872, China} 

\date{\today}
\begin{abstract}
We investigate the exact nature of the superfluid--to--Mott--insulator 
crossover for interacting bosons on an optical lattice in a one--dimensional, 
harmonic trap by high--precision density--matrix renormalization--group 
calculations. The results reveal an intermediate regime characterized by a 
cascade of microscopic steps. These arise as a consequence of individual 
boson ``squeezing'' events and display an even--odd alternation dependent 
on the trap symmetry. We discuss the experimental observation of this 
behavior, which is generic in an external trapping potential.
\end{abstract}
\maketitle

The recent rapid developments in ultra--cold--atom experiments for both 
bosons and fermions have greatly stimulated the fuller exploration of a 
number of fundamental properties of strongly 
correlated systems \cite{rbdz}. One of the most important phenomena to 
be observed and characterized is the transition from a superfluid to a 
Mott--insulator phase displayed by interacting bosonic atoms in optical 
lattices \cite{greiner02}. The Bose--Hubbard model \cite{Fisher89} shares 
a number of the properties of the Hubbard model for correlated electrons, 
whose properties, including superconductivity and Mott--insulating behavior, 
have challenged condensed matter physicists for half a century. This model 
can be realized in one, two, and three dimensions on an optical lattice 
\cite{Jaksch98}, with the enormous advantage over electronic solids that 
the ratio between kinetic and interaction parameters of the particles is 
in principle continuously tunable. 

In experiments, a trapping potential is required to confine the atoms. 
It was shown both theoretically \cite{white-boson} and experimentally 
\cite{greiner02,stoeferle04,gerbier05} that the pure Mott--insulating 
phase, where the boson distribution is uniform despite the trap profile, 
is obtained only very at high interaction strengths. Otherwise, a 
spatial ``shell structure'' of the local density is found \cite{Jaksch98,
foelling06,sci06}, with Mott--insulating and superfluid regions  
present simultaneously in regions of different trap depth. Extensive 
studies of this behavior include a demonstrated compression of the 
superfluid region as the Mott insulator is approached \cite{gerbier06} 
and quantitative calculations both of the shell structure \cite{batrouni02, 
kollath04} and of accompanying features in the visibility \cite{sengupta05}.

Here we show that the nature of the Mott transition in a trapping 
potential is considerably more complex than recognized previously: 
``superfluid'' bosons are expelled one by one into the Mott--insulating 
regions with increasing interaction strength. The intermediate regime where 
this sequence of microscopic quantum transitions takes place is the necessary 
consequence of the additional 
energy scale introduced by the trap, and the resulting physics is 
qualitatively different from the infinite system.

\begin{figure*}[t]
\includegraphics[width=4.2cm]{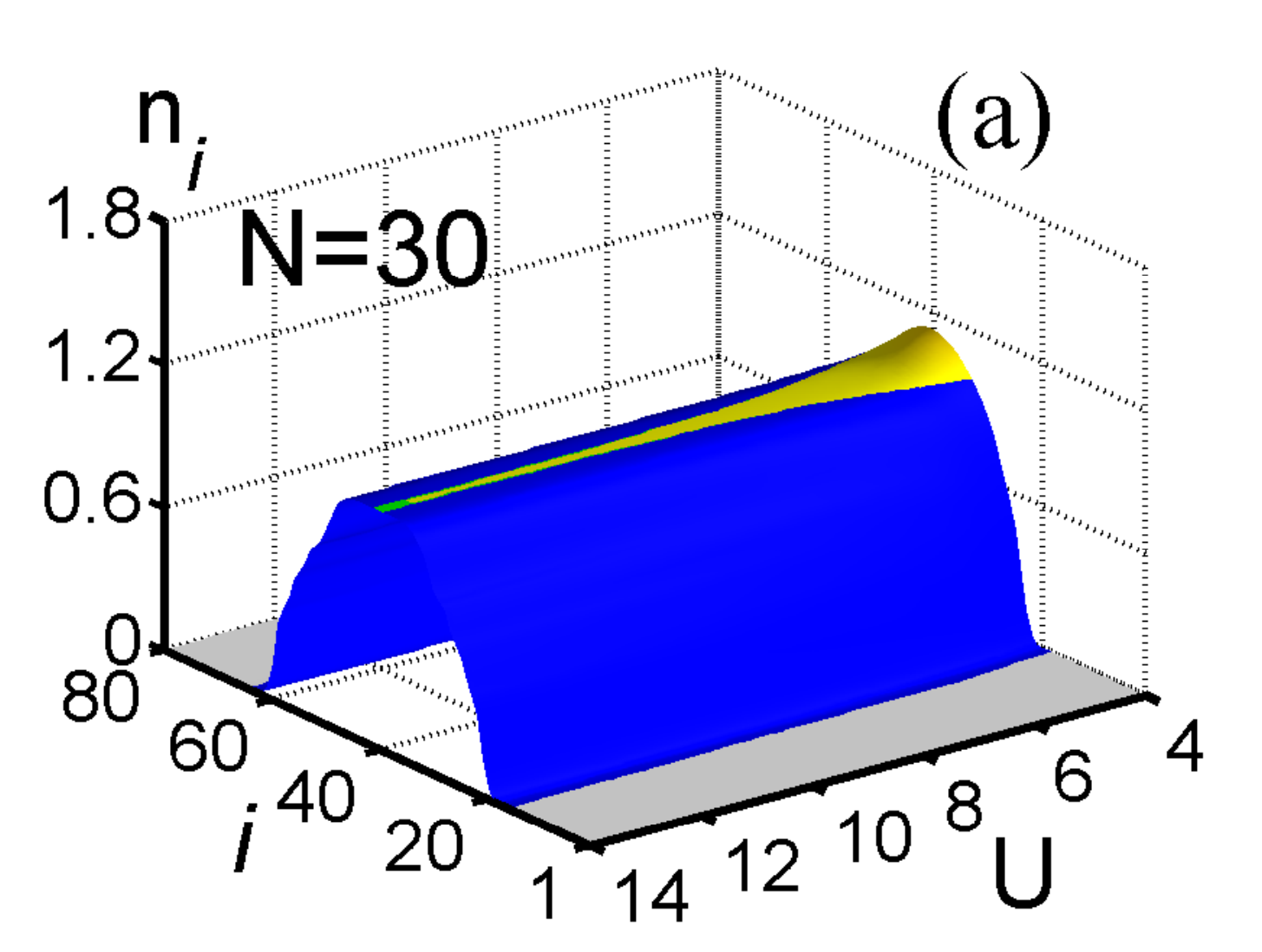}\includegraphics[width=4.2cm]{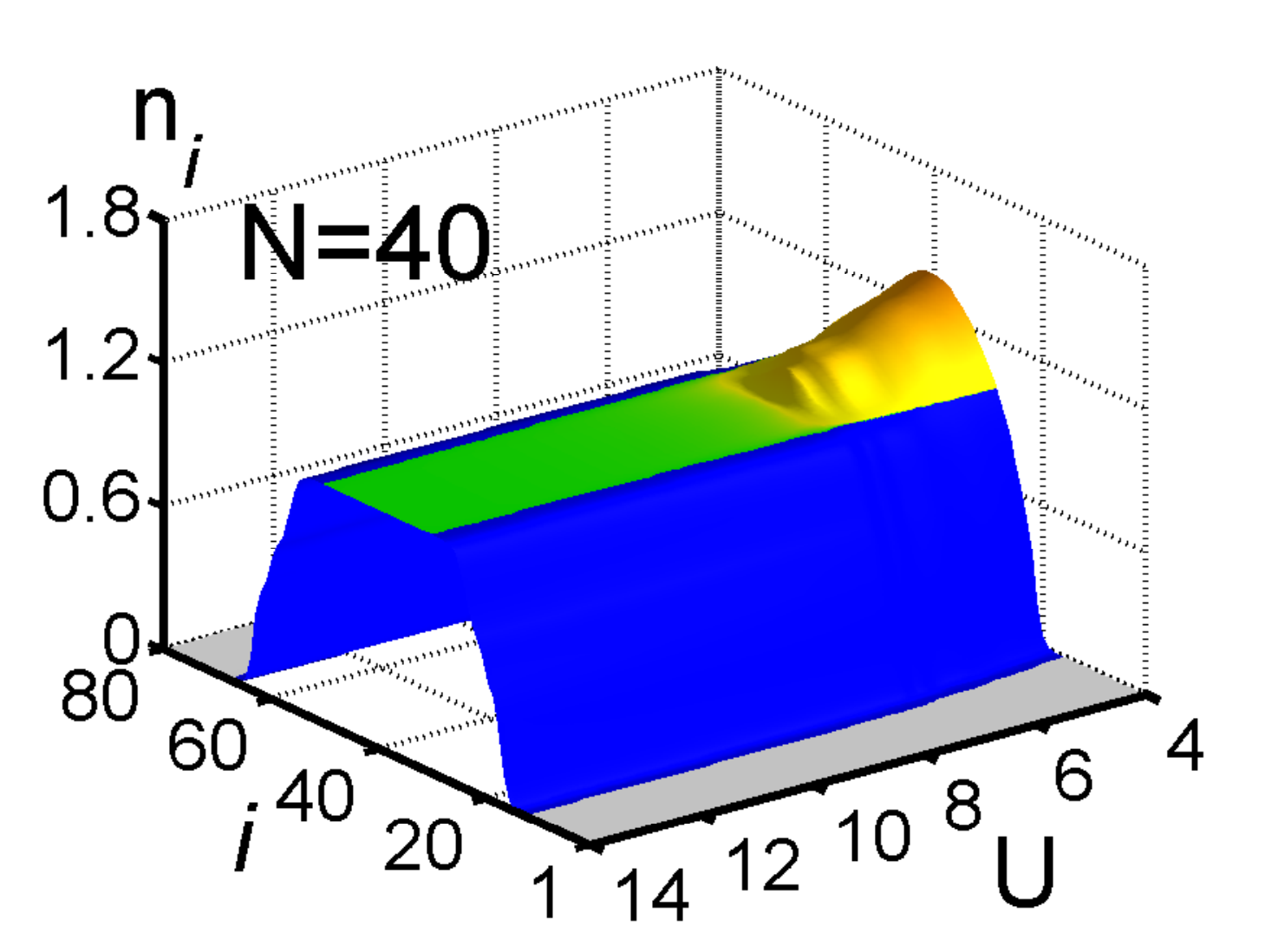}\includegraphics[width=4.2cm]{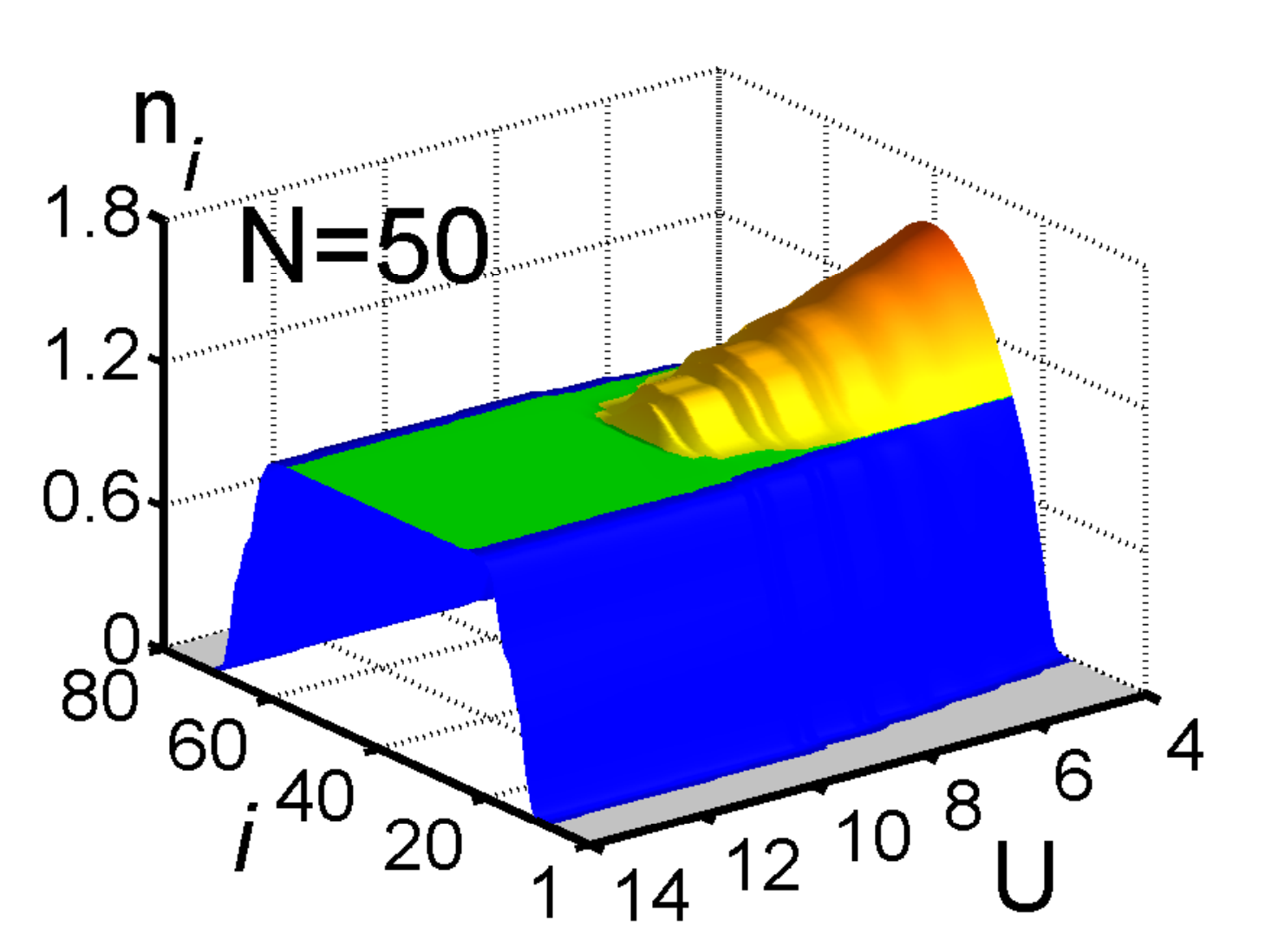}\includegraphics[width=4.2cm]{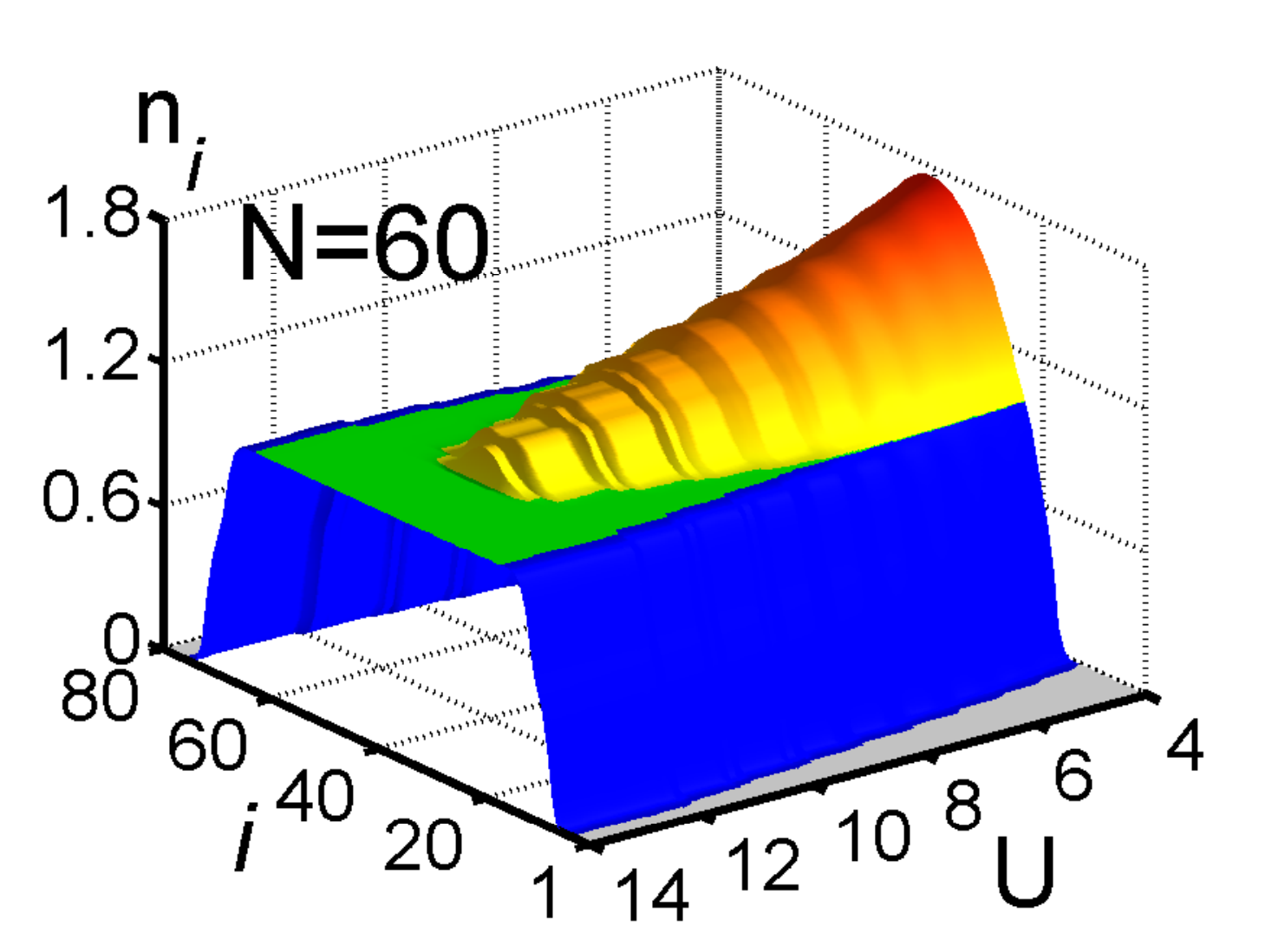}
\includegraphics[width=4.44cm]{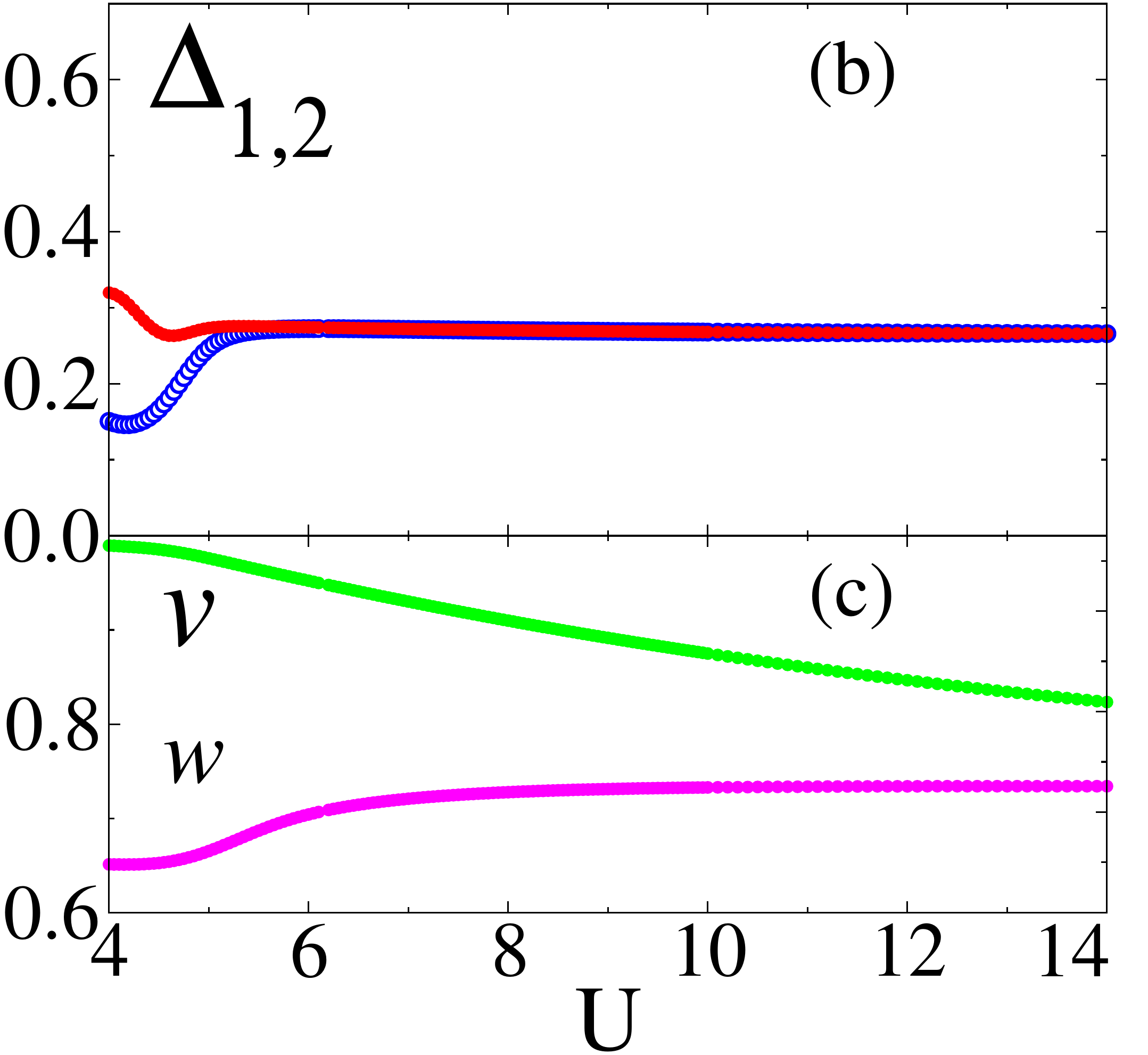}\includegraphics[width=4.12cm]{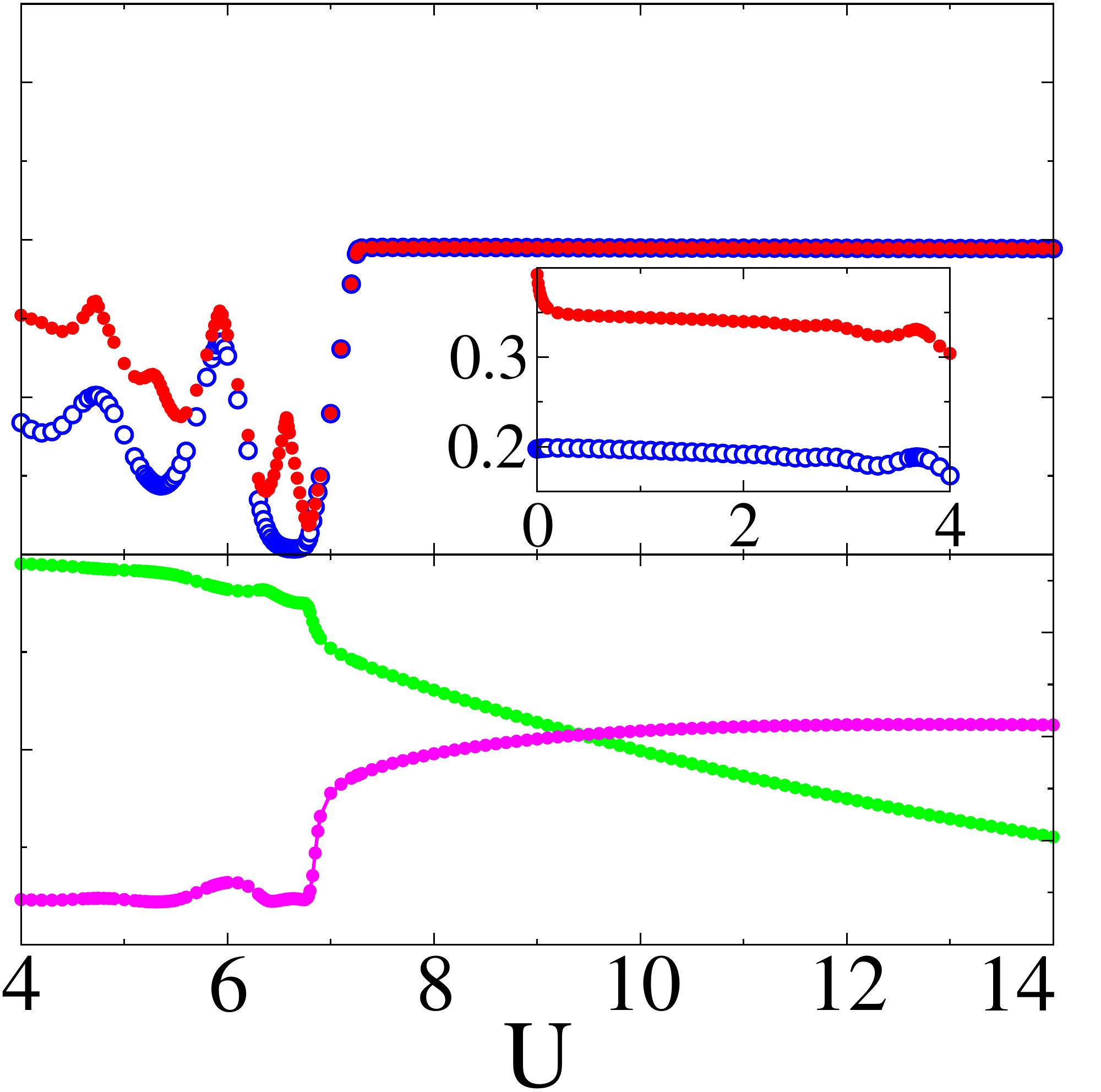}\includegraphics[width=4.1cm]{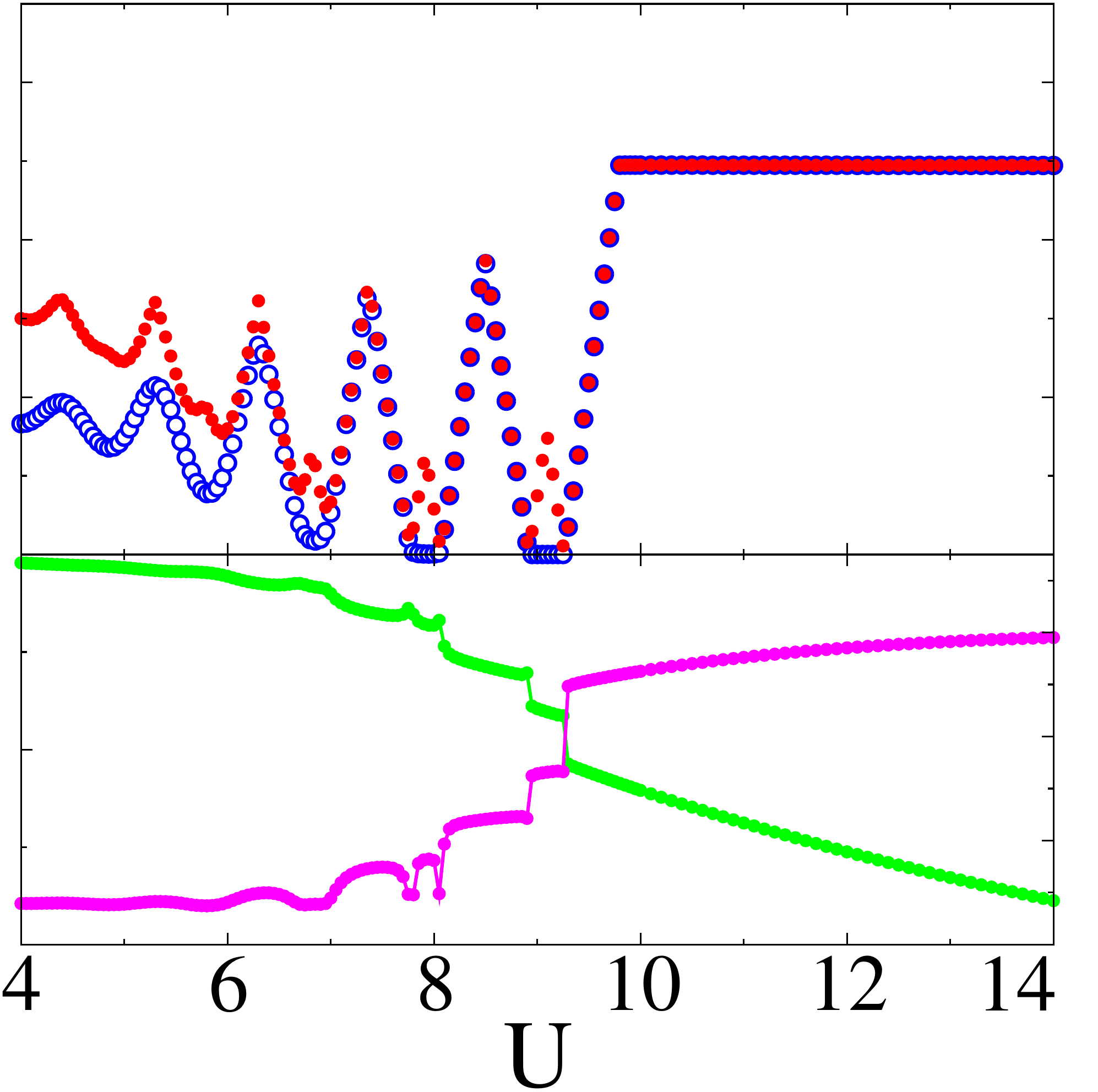}\includegraphics[width=4.37cm]{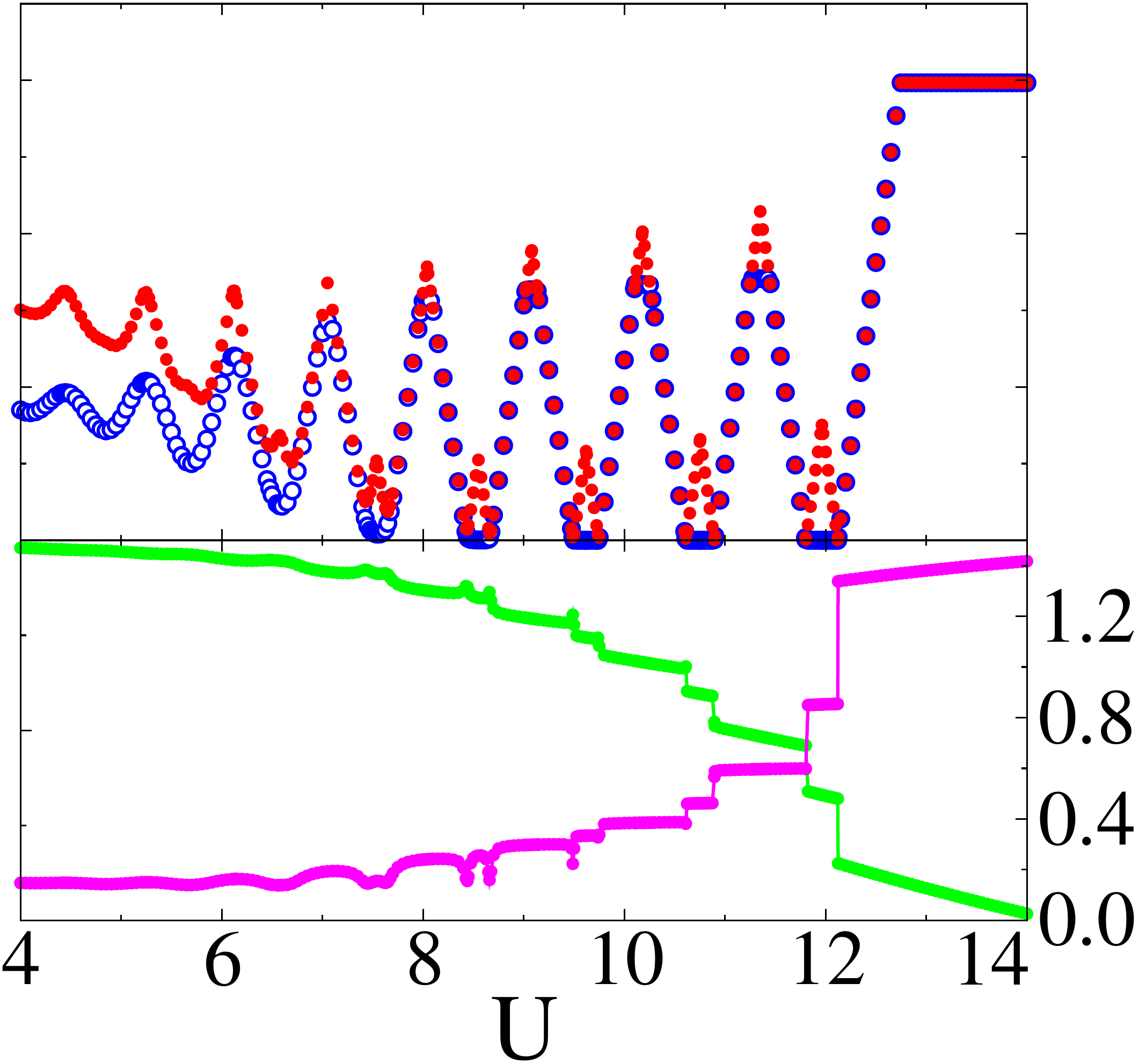}
\caption{(color online). Results for $V_T = 0.01$, $t = 1$, and system length 
$L = 80$ for $N = 30$, $40$, $50$, and $60$ particles. (a) Density profiles 
$n_i$ as functions of $U$. (b) Energy gaps $\Delta_{1,2}$; inset: 
$\Delta_{1,2}$ for low $U$. (c) Visibility $\nu$ (left axis) and peak width 
$w$ (right).}
\label{LDDU}
\end{figure*}

To explore the nature of the states intermediate between the perfect 
superfluid and the Mott insulator in a trap, we consider the one--dimensional 
(1D) Bose--Hubbard model with a harmonic trapping potential. This system is 
realized experimentally for a cloud of interacting bosons loaded in an 
optical lattice with one dominant recoil energy. The Hamiltonian is 
expressed \cite{Fisher89, Jaksch98} as
\begin{eqnarray}\label{hubham}
\hat H & = & - t \!\! \sum_{i=1,L} \!\! \left(\hat a^{\dagger}_i \hat a_{i+1}
 + \hat a^{\dagger}_{i+1} \hat a_i\right) + {\textstyle \frac12} U \!\! 
\sum_{i=1,L} \!\! \hat n_i (\hat n_i-1) \nonumber \\ & & + V_T \!\! 
\sum_{i=1,L} \!\! \hat n_i [i-(L+1)/2]^2,
\end{eqnarray}
where the lattice spacing is set to 1 and the on--site interaction $U$ and 
(bond--centered) trapping potential $V_T$ are measured in units of the 
hopping integral $t$. We apply the density--matrix renormalization--group 
(DMRG) technique \cite{rdmrg}, using a specially modified finite--system 
algorithm optimized for the accuracy of the low--lying energy states. 
An appropriate sweeping procedure is essential to guarantee 
convergence in this delicate problem \cite{rxwk}. Sweeping is conducted 
gradually from the middle to the two ends, convergence is obtained for each 
sweeping length $l \in [4,L]$, and the three lowest eigenstates are targeted 
simultaneously. Extremely high accuracies are required to allocate reliably 
the symmetries of degenerate states, and these were examined \cite{rhwwyt} 
for different system sizes $L$, numbers $n$ of bare states per site, and 
numbers $m$ of states retained per block, for each total boson number $N$ 
and interaction $U$. Working up to $n = 32$ and $m = 1200$, we found that 
$n = 8$ and $m = 100$ are sufficiently large for $U \in [4,14]$ to ensure 
truncation errors smaller than the symbol sizes in the figures below, while 
$n = 32$ and $m = 400$ are necessary for $U < 4$. 

Systematic studies were performed for systems with $20 \le N \le 70$ 
bosons in a trap with $V_T = 0.01$, for which $L = 80$ is sufficient 
to remove any boundary effects. The unconventional properties of the 
intermediate regime begin at $N = 40$ and become progressively clearer 
with increasing $N$ until a higher Mott plateau is reached. We have 
performed many calculations to amplify the current results \cite{rhwwyt}, 
but this parameter and filling regime, shown in Fig.~1, ensures a minimal 
system reflecting all of the qualitative physics. We focus only on the 
ground state and not on the nature of the excitations 
\cite{rhwwyt}; however, because $U$--induced changes in the ground state 
appear as (avoided) level--crossings, it is instructive to consider the 
lowest three energy levels, $E_{0,1,2}$, represented in Fig.~1 
by the gaps $\Delta_{1,2} = E_{1,2} - E_0$. We stress again that this 
is a finite system, both theoretically and in experiment: we provide 
numerically exact results for the intrinsic physics of this system, and 
there are no ``finite--size effects'' in the conventional sense of 
approximations to a thermodynamic limit. We also calculate the particle 
numbers ($n_i^{s,d,t}$) for single, double, and triple occupation, the 
visibility $\nu$, and the peak height $S_{\rm max}$ and width $w$ (taken 
as the full width at half the sum of the maximum and minimum values) of 
the momentum distribution function. The measurable physical quantities 
$\nu$ and $w$ \cite{rbdz} illustrate most clearly the effects we consider
here.

Figure 1 shows three clear regimes. At low $U$ is the bell--shaped 
distribution of a ``superfluid'' (SF) system, which has strong visibility 
and a trap--related gap above the ground--state condensate. At high $U$ 
is the flat distribution ($n_i = 1$), despite the energy cost of the 
trap at the edges, of a pure ``Mott insulator'' (MI) regime, with falling 
visibility and gaps of order $t$ to states at the edges. Between these two 
limits is a complex intermediate phase characterized by oscillating and 
vanishing energy gaps, step--like changes in the particle distribution and 
the visibility, and a spatial ``shell structure'' with MI plateaus around 
a central SF region. Note that we use the abbreviations SF and MI to refer 
not to the bulk phases but respectively to the local regions of the 
distribution shown by the red--yellow curve and by the green plateaus. 
The almost constant shape of the blue ``tails'' in $n_i$ at the system 
edges is dictated not by $U$ but by $V_T$.

For the homogenous Bose--Hubbard model, the Mott transition occurs at $U_c 
\! = \! 3.61 \! \pm \! 0.13$ \cite{white-boson}. In a trap, there is little 
evidence of the energy scale $U_c$, although it corresponds loosely to the 
onset of oscillations in $\Delta_{1,2}$ and in $n_i$. It is easy to deduce 
that a pure MI phase is established only when the energy cost $U$ of the 
last boson in the central SF region exceeds that of placing it at the 
edge. One effective description of this process is to write $U_{c2}
 - \alpha(1) t \simeq {\textstyle \frac{1}{4}} V_T (N - \delta)^2$, where 
$\alpha(1) \sim 4$ \cite{rhwwyt} represents a kinetic--energy contribution 
due to the last boson and $\delta \sim 6$ [Fig.~2(b)] represents the 
edge effects at lowest order. Both $\alpha$ and $\delta$ are corrections 
to a leading quadratic dependence of $U_{c2}$ on $N$, arising directly 
from the harmonic trapping potential, which makes the width in $U$ of 
the intermediate regime very significant. This illustrates directly how 
$V_T$ acts as an additional energy scale, which is responsible for the 
presence of the additional, intermediate regime. Clearly, if $V_T = 0$ 
the atoms are no longer trapped, or their density controlled: the trap 
($V_T$) is a non--perturbative term in ${\hat H}$ (\ref{hubham}), ensuring 
that any ensemble of trapped atoms is generically in a different class from 
the homogeneous, infinite system.

\begin{figure}[t]
\includegraphics[width=4.3cm]{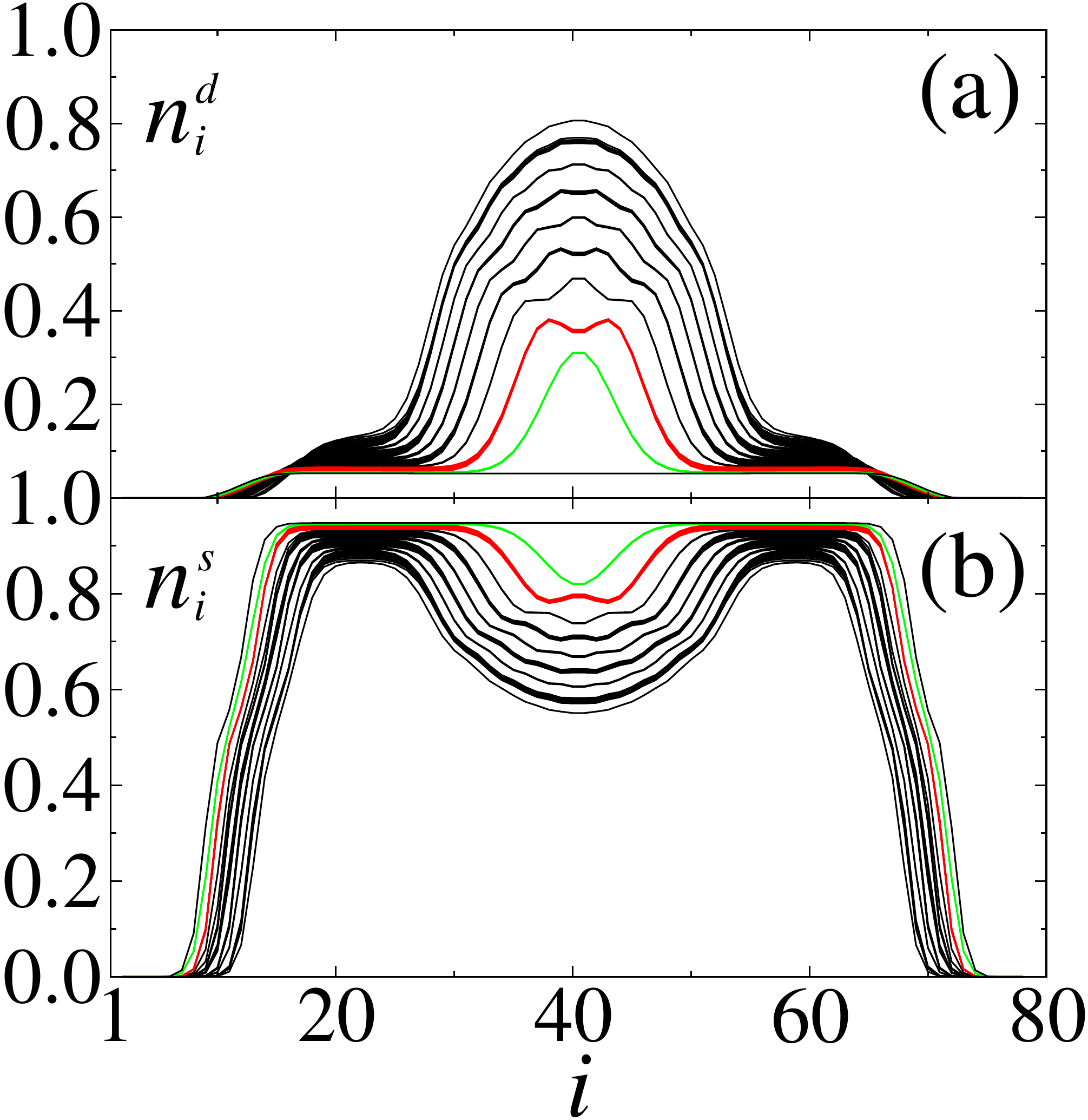}\includegraphics[width=4.0cm]{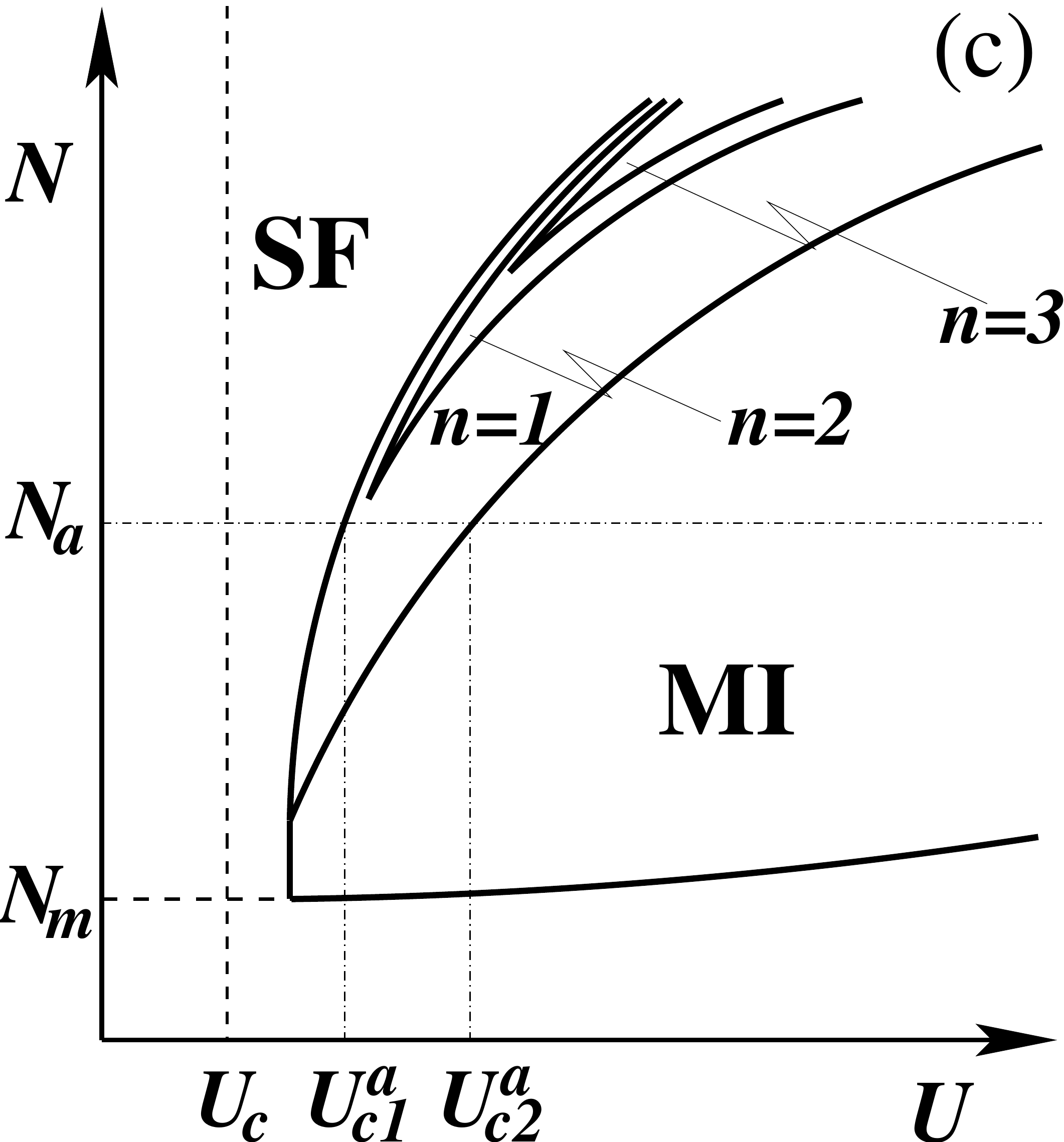}
\caption{(color online). Superposed distributions of particles in (a) doubly 
occupied sites ($n^d_i$) and (b) singly occupied sites ($n^s_i$) for $N = 60$, 
$L = 80$, and $V_T = 0.01$. Shown in each panel are 47 curves for values $7.6 
\le U \le 12.2$, which fall into 10 different sets corresponding to $0, \dots 
9$ bosons in the central SF region. Red curves for two remaining SF bosons 
and green for one. (c) Schematic representation of generic ``state diagram'' 
for a trapped boson system as a function of $U$ and $N$. Solid lines mark 
finite--system ``phase transitions,'' dot--dashed lines a typical filling 
$N_a$ with two $n \! = \! 1$ Mott plateaus.} 
\label{n1n2}
\end{figure}

As shown in Fig.~1, the width in $U$ of the intermediate regime indeed 
increases strongly with $N$. The oscillations in $\Delta_{1,2}$ and $n_i$ 
grow with $U$ until level--crossings and steps appear (we refer to 
``plateaus'' in $i$ and ``steps'' in $U$). In this cascade of steps, 
$\Delta_1$ vanishes over short but finite ranges of $U$, indicating the 
double degeneracy of the ground state when the two MI plateaus contain 
an odd number of particles. Each level--crossing represents a squeezing 
process where one particle is ejected from the central SF region and 
placed on the outer edge of one or other plateau: these processes are 
quantized. The locations and number of these crossings, which correspond 
directly to jumps in $n_i^d$, $\nu$, and $w$ as functions of $U$, are not 
universal and nor are their spacings identical: the examples in Fig.~1 favor 
even numbers of particles in the central SF region (``even--odd effect''). 
We note that such series of discrete, single--particle transitions are 
present in Monte Carlo simulations performed for a harmonic trap with 
$n \! = \! 1$ and $n \! = \! 2$ MI plateaus \cite{sengupta05} and in 
DMRG calculations for a double--well trap \cite{rjwx}; however, neither 
set of authors appeared to recognize their significance as quantized 
squeezing, or noticed an even--odd effect. 

Additional insight into the squeezing process can be found in Figs.~2(a) 
and (b), which show the evolution with $U$ of $n_i$ across the intermediate 
regime for $N = 60$ (right panels of Fig.~1). While $n_i^t$ and higher site 
occupations are always negligible, $n_i^s$ and $n_i^d$ are perfectly 
anticorrelated. A key qualitative point must be made here: although $n_i = 
1$ in the MI plateaus, this is not due to perfect single--site occupation 
(because $n_i^s \ne 1$), and the finite constant $n_i^d$ reflects the 
complete coherence of the many--body wave function mediated by continuous 
kinetic processes coupling the central SF to the edges through the MI 
regions. In spite of this system--wide coherence, $n_i$ does not evolve 
continuously, but instead undergoes 9 abrupt changes across the 
intermediate regime [Figs.~2(a) and (b)] as $U$ acts to squeeze the 
initial 9 bosons in the central SF region successively down to zero. 
The areas represent the exact boson numbers, giving complete quantitative 
information on the spatial extent and the structure of the central SF part 
of the wave function, the number of weak maxima for 2, 3, $\dots$ particles 
reflecting the boson phase coherence. 

\begin{figure}[t]
\includegraphics[width=4.39cm]{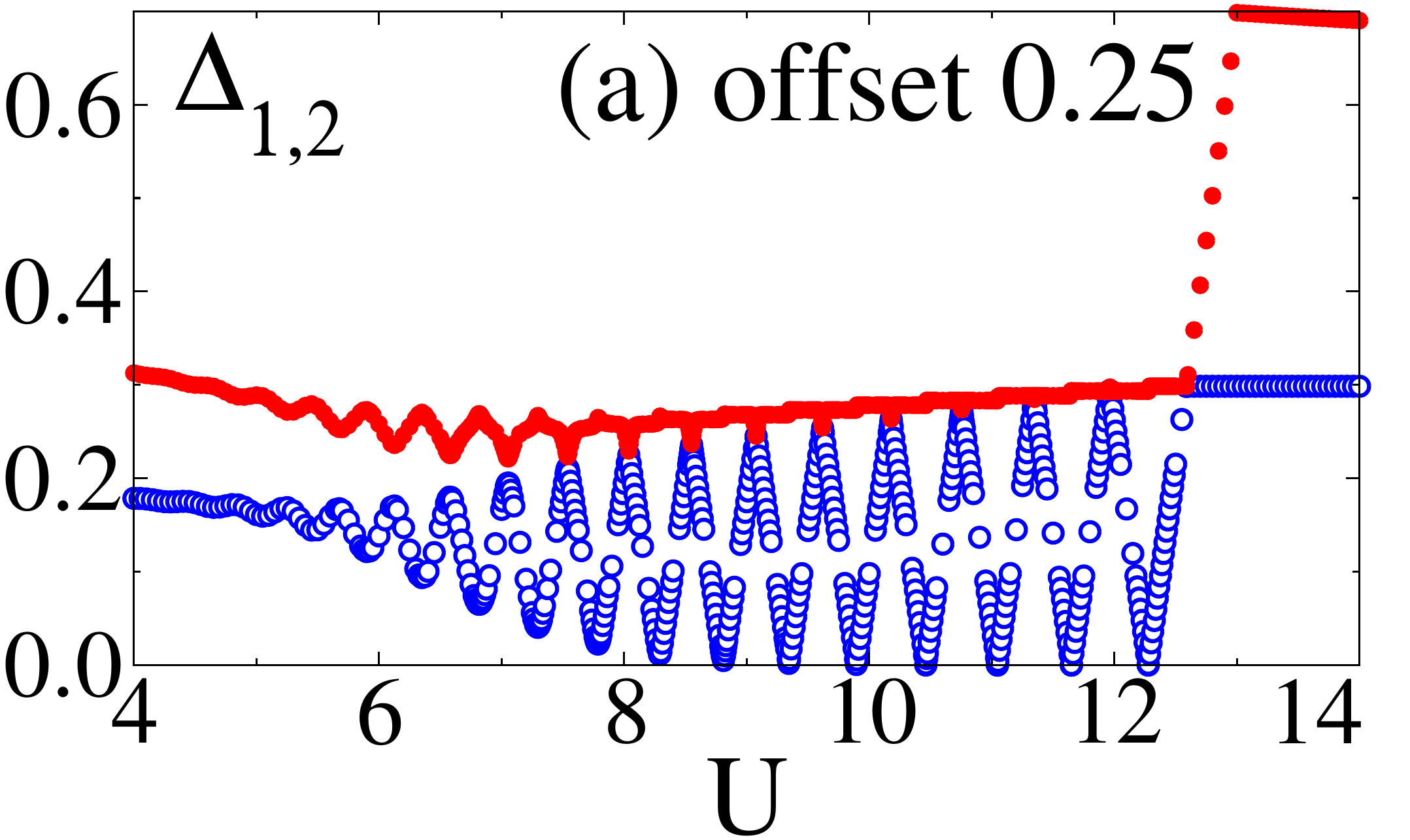}\includegraphics[width=4.1cm]{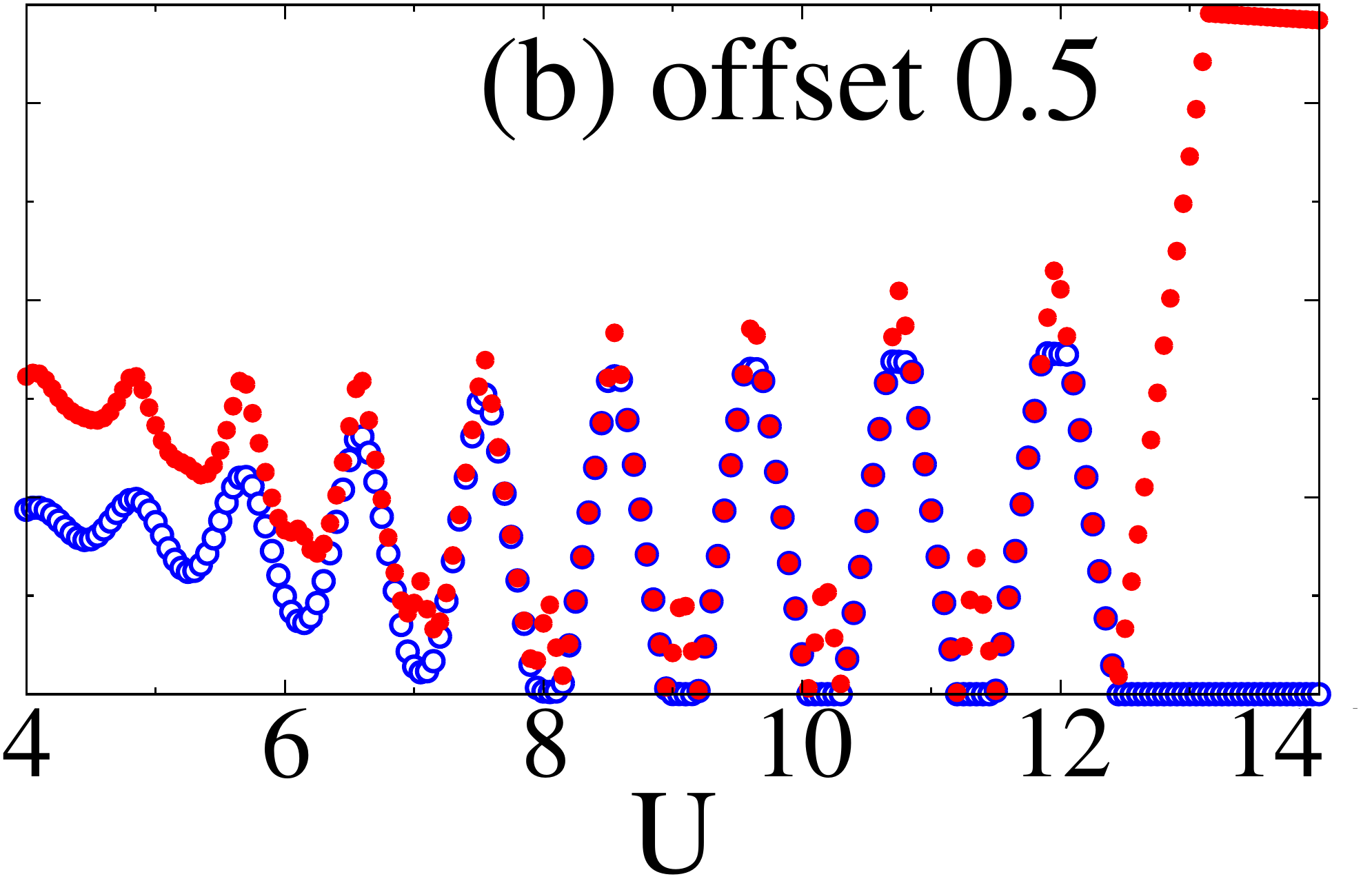}
\caption{(color online) Lowest energy gaps for $N = 60$, $L = 80$, and 
$V_T = 0.01$ with trap center offsets 0.25 (a) and 0.5 (b). The case with 
offset 0 is shown in Fig.~1 (center right panel).}
\label{fig3}
\end{figure}

Returning to the quantitative information in Fig.~1(b), and to the question 
of a ``phase diagram'' for the finite system of trapped bosons, we define 
the onset $U_{c1}$ of an intermediate phase, which corresponds to the 
appearance of two true MI plateaus as opposed to flattening shoulders, from 
the first level--crossing. Proceeding as above, $U_{c1} - \alpha (N_{\rm 
max}^{\rm sf} \! + \! 1) t \simeq {\textstyle \frac{1}{4}} V_T (N - \delta
 - N_{\rm max}^{\rm sf})^2$, where $N_{\rm max}^{\rm sf}$ is the number of 
bosons accommodated in the central SF region at $U_{c1}$ and $- \alpha (N_{\rm 
max}^{\rm sf} \! + \! 1) t$ represents the kinetic energy of the squeezed 
boson. The leading dependence of the width of the intermediate regime is 
then $U_{c2} - U_{c1} \sim {\textstyle \frac{1}{2}} V_T N N_{\rm max}^{\rm 
sf}$, as shown schematically in Fig.~2(c). This type of ``canonical phase 
diagram,'' or ``state diagram'' \cite{batrouni02}, is appropriate for 
cold--atom experiments where a fixed number of particles is loaded and 
the optical lattice tuned adiabatically. Figure 2(c) also shows that no 
MI can be formed if $N$ is too low ($N < N_m$), while for $N$ sufficiently 
large, higher plateaus will appear. Although the functional forms and 
relative positions of the phase boundaries can be changed by altering 
the trap shape, this is the generic state diagram of any ensemble of 
bosons whose spatial distribution is contained by a finite external 
potential.

A further essential qualitative feature of single--boson squeezing is its 
even--odd asymmetry. Figure 1 shows even boson numbers $N$ in a bond--centered 
(``even'') trap, where the steps of even boson occupation in the central SF 
region are wider (more stable) than the odd ones. However, if $N$ is odd, 
the situation is exactly reversed, and the odd--occuption SF steps are 
wider \cite{rhwwyt}. To investigate this effect systematically, we move 
the trap midpoint continuously from the bond--centered position to a 
site--centered one (``odd trap''). Again the situation is reversed in the 
site--centered limit (offset 0.5), with odd steps more stable [Fig.~3(b)] 
for even $N$. This suggests a very straightforward interpretation: as $U$ 
is increased, two states, with the squeezed boson at one side of the system 
or the other, simultaneously become more favorable than the states at the 
center. This picture is exact in the $t \rightarrow 0$ limit, where it is 
clear analytically that bosons are squeezed out of the central SF region 
in pairs, while the effect of a finite kinetic--energy term $t$ is to 
stabilize a narrow (in $U$) step of the intermediate, symmetry--disfavored 
boson number. Whether even or odd steps are favored is a simple consequence 
of the trap symmetry and filling. The confirmation that for spinless bosons 
there are no special, pairwise correlation effects in the SF or MI regions 
is obtained by following the evolution of the asymmetry with the trap offset: 
specifically, there is no even--odd effect at all for offset 0.25 [Fig.~3(a)], 
confirming its origin in the trapping potential.

Having understood the physics of quantized squeezing, it is possible 
to comment on its manifestations in a less minimal context. The same 
squeezing effect clearly operates with more bosons in the trap, within 
and between more complex shell structures of higher MI plateaus. We also 
find an even--odd effect when squeezing bosons from central SF or MI 
regions to the $n \! = \! 2$ or higher Mott plateaus, and not only to 
the system edge (the $n \! = \! 1$ plateau). Quantized squeezing 
processes occur in all dimensions as individual particles within the 
coherent, many--body wave function are pushed successively out of the 
central SF region with increasing $U$. In dimensions greater than one, much 
more elaborate even--odd effects would be expected: as one example, the 
generic state degeneracy for a square lattice in a circular trap (2D) is 
8, and thus cascades of seven narrow steps might be expected between wide 
8th steps for a centered trap. For a cubic lattice and spherical trap (3D), 
this number is 48. A more challenging question in the physics of strongly 
correlated systems is posed by the situation with finite--spin bosons or 
several boson species in the trap, where interaction effects may lead to 
intrinsic pairing (or more complex $n$--merization) in the SF, or indeed 
the MI, region. This would generate a completely separate mechanism for 
the tendency of particles to move from SF to MI regimes in groups rather 
than singly.

\begin{figure}[t]
\includegraphics[width=4.2cm]{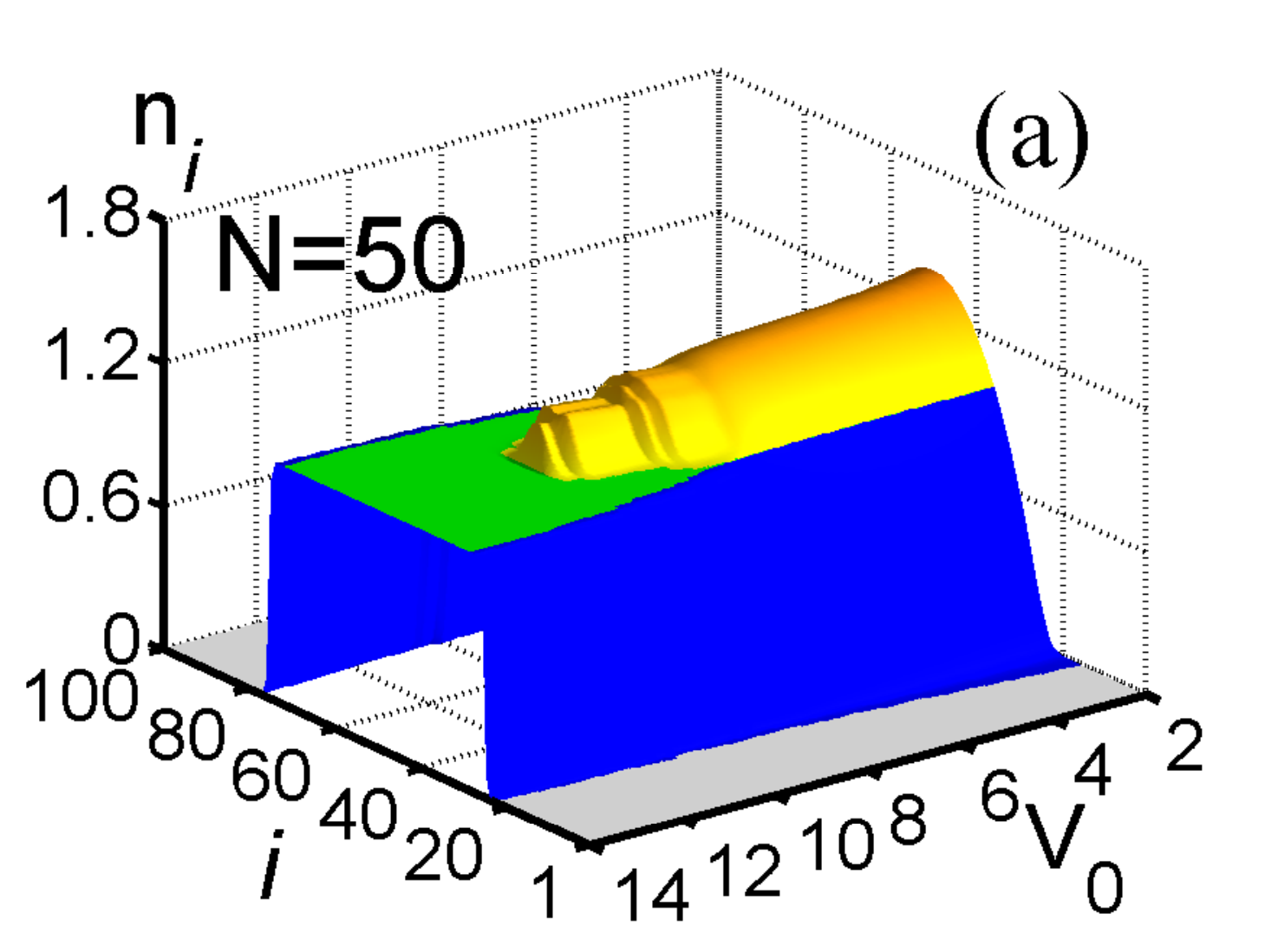}\includegraphics[width=4.2cm]{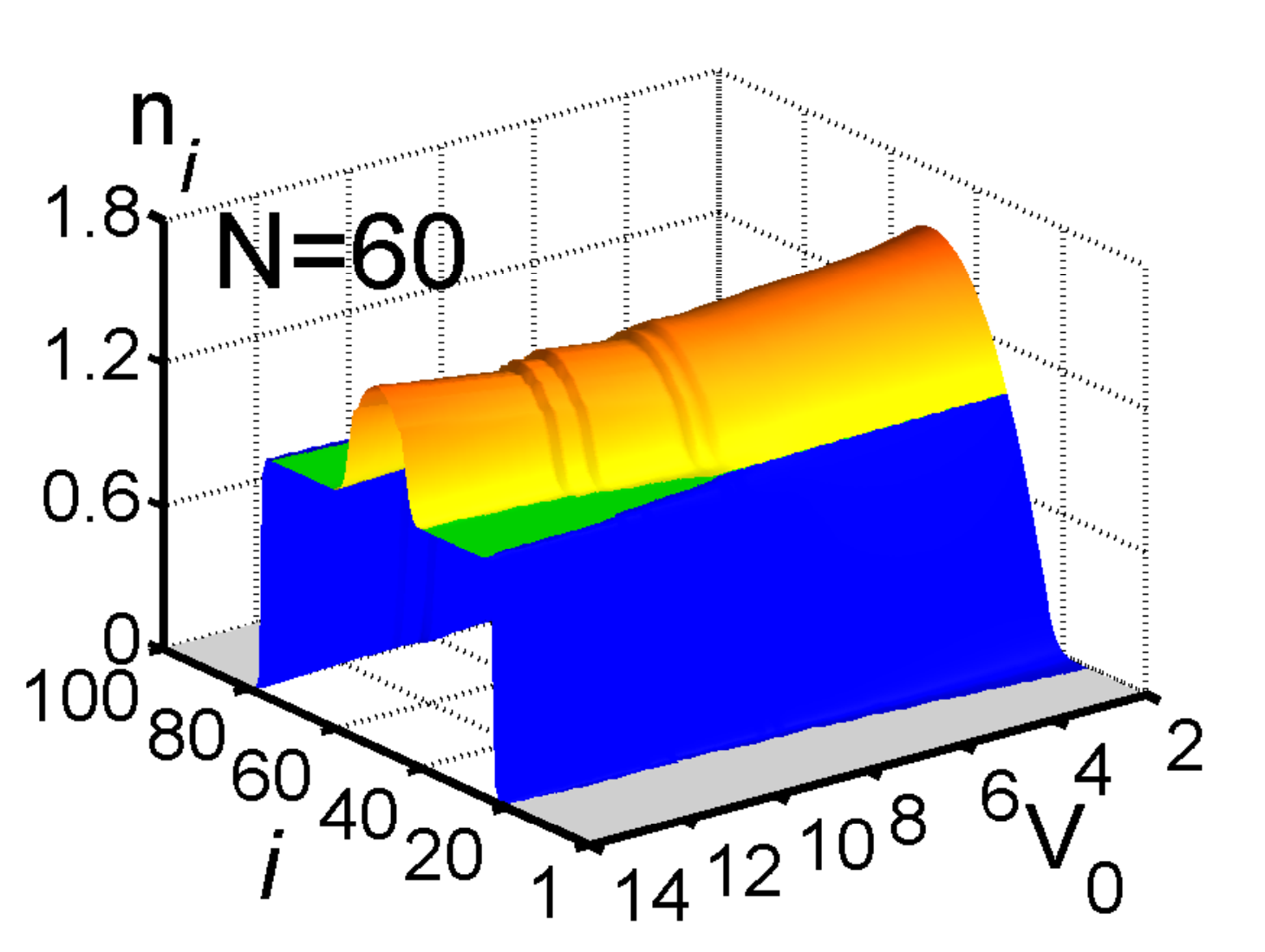}
\includegraphics[width=4.19cm]{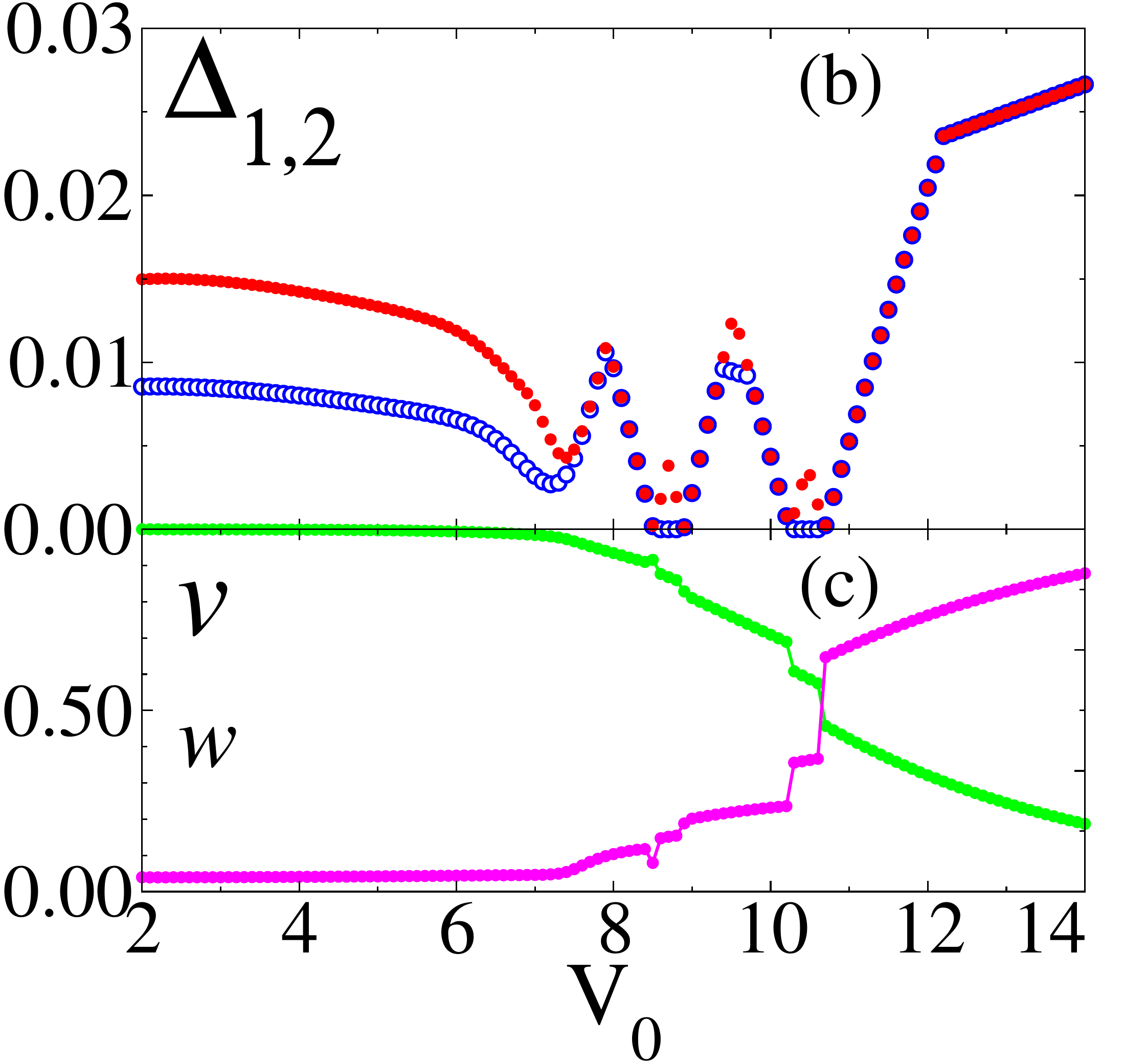}\includegraphics[width=4.0cm]{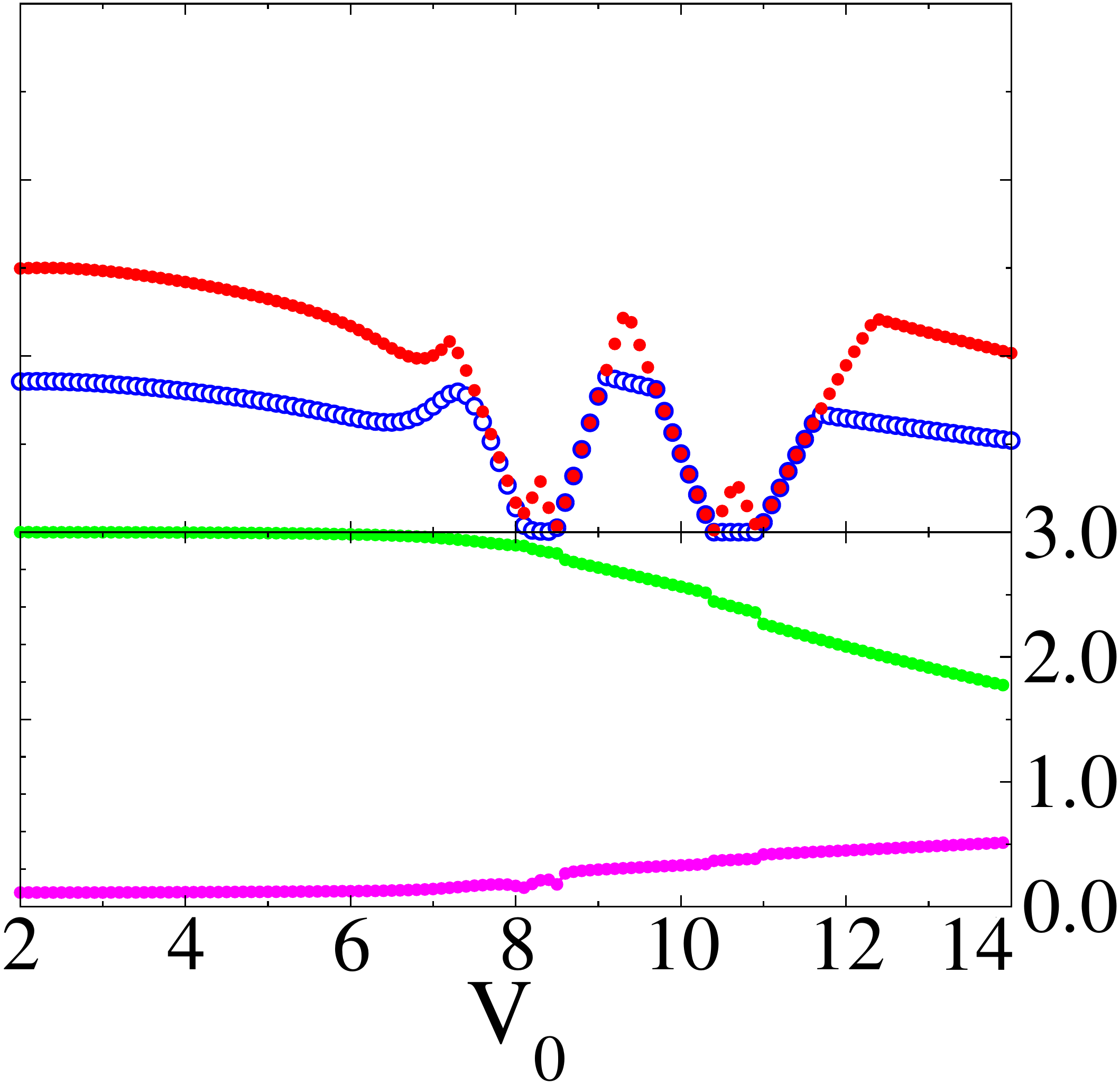} 
\caption{(color online) Results shown in the format of Fig.~1 for $N = 50$ 
and 60 particles in a 100--site trap as a function of the optical lattice 
depth parameter $V_0$ for $E_r = 1$ (see text).}
\label{fig4}
\end{figure}

We close by discussing how quantized squeezing and even--odd asymmetries 
would be observed in experiment, restricting our considerations to 1D. 
Experimental measurements on optical lattices are performed by varying the 
lattice depth $V_0$ in units of the recoil energy $E_r$. Following the 
calculations of Ref.~\cite{gerbier05}, we have taken $t/E_r = 1.43 
(V_0/E_r)^{0.98} \exp[-2.07 \sqrt{V_0/E_r}]$, $U/E_r = 0.0386(V_0/E_r) 
^{0.88}$, and $V_T/E_r = (5.332 + 3.427 V_0/E_r) \times 10^{-5}$. 
Clearly, varying $V_0/E_r$ explores one curve in the 2D space of the 
parameters $U$ and $V_T$ for each $N$. Figure 4 shows the numerical 
results obtained as a function of $V_0$ for $N = 50$ and $60$. It remains 
possible, but not simple, to find parameters and trap fillings for which 
all three regimes can be observed, although this requires scanning over 
a rather broad range of $V_0$. While the boson--squeezing plateaus are 
widely spaced in $V_0$, their even--odd asymmetry becomes very pronounced. 
As a final caveat, some concern has been raised recently \cite{rpkvt} 
over the heating effects occurring on changing the lattice depth for 
pre--trapped bosons due to energy absorption from the optical field, 
and appropriate measures are required to remain sufficiently close to 
the adiabatic limit.

In summary, we show that there exists a regime intermediate between the 
pure superfluid and the Mott--insulator phase for interacting bosons in 
a one--dimensional optical lattice with a harmonic trapping potential. 
We have investigated the sequence of microscopic quantum transitions 
across this regime. These processes reflect the quantized squeezing of 
bosons into the Mott--insulator region, and occur as discrete events 
despite the overall coherence of the many--body wave function. The 
even--odd asymmetry of the squeezing phenomenon is a direct 
consequence of the trap symmetry. The ``phase diagram'' of the trapped 
system is intrinsically, qualitatively different from the infinite 
system, the intermediate regime being the necessary consequence of 
the additional energy scale introduced by the trap. The effects 
revealed by detailed studies of this sort mandate a broader 
interpretation of the ``Mott transition'' in cold--atom systems. 

We are grateful to T. Xiang, Lu Yu, F. C. Zhang, and J. Z. Zhao for
fruitful discussions. This work was supported in part by grants
NFC2005CB32170X and C10674142.


\begin{thebibliography}{99}

\bibitem{rbdz} 
I. Bloch, J. Dalibard, and W. Zwerger, Rev. Mod. Phys. {\bf 80}, 885 (2008).

\bibitem{greiner02}
M. Greiner {\it et al.}, Nature (London) {\bf 415}, 39 (2002).

\bibitem{Fisher89} M. P. A. Fisher, P. B. Weichman, G. Grinstein, and 
D. S. Fisher, Phys. Rev. B {\bf 40}, 546 (1989).

\bibitem{Jaksch98} D. Jaksch {\it et al.}, Phys. Rev. Lett. {\bf 81}, 
3108 (1998).

\bibitem{white-boson} T. K\"{u}hner, S. White, and H. Monien, Phys. Rev. 
B \textbf{61}, 12474 (2000).

\bibitem{stoeferle04} T. St\"{o}ferle {\it et al.}, Phys. Rev. Lett. 
{\bf 92}, 130403 (2004).

\bibitem{gerbier05} F. Gerbier {\it et al.}, Phys. Rev. Lett. {\bf 95}, 
050404 (2005).

\bibitem{foelling06} S. F\"{o}lling {\it et al.}, Phys. Rev. Lett. {\bf 97}, 
060403 (2006).

\bibitem{sci06} G. K. Campbell {\it et al.}, Science {\bf 313}, 649 (2006).

\bibitem{gerbier06} F. Gerbier {\it et al.}, Phys. Rev. Lett. \textbf{96}, 
090401 (2006).

\bibitem{batrouni02} G. G. Batrouni {\it et al.}, Phys. Rev. Lett. {\bf 89}, 
117203 (2002).

\bibitem{kollath04} C. Kollath, U. Schollw\"{o}ck, J. von Delft, and 
W. Zwerger, Phys. Rev. A \textbf{69}, 031601 (2004).

\bibitem{sengupta05} P. Sengupta {\it et al.}, Phys. Rev. Lett. \textbf{95}, 
220402 (2005).

\bibitem{rdmrg} I. Peschel, X. Q. Wang, M. Kaulke, and K. Hallberg,
{\sl Density Matrix Renormalization}, (Springer, Berlin, 1999); 
U. Schollw\"ock, Rev. Mod. Phys. {\bf 77}, 259 (2005).

\bibitem{rxwk} 
X. Wang, Mod. Phys. Lett. B {\bf 12}, 667 (1998).

\bibitem{rhwwyt} 
S. Hu {\it et al.}, unpublished.

\bibitem{rjwx} 
H. C. Jiang, Z. Y. Weng, and T. Xiang, Phys. Rev. B {\bf 76}, 224515 (2007). 

\bibitem{rpkvt} 
L. Pollet, C. Kollath, K. Van Houcke, and M. Troyer, New J. Phys. 
{\bf 10}, 065001 (2008).

\end{thebibliography}
\end{document}